# Triaxial Asymmetry Driven Rotational Dynamics and Lateral Equilibrium Position in Inertial Flow

Takayuki Suzuki[1], Anna B. Stephenson[2,3], Jinsik Yoon[4], Junghyun Bae[5], Sung-Eun Choi[6], Kilho Son[7], Diego Alba Burbano[1], Harrison Khoo[1], Wook Park[8], and Soojung Claire Hur[1,9-12*]

[1]Department of Mechanical Engineering, Johns Hopkins University, Baltimore, Maryland, 21218 USA
[2]Department of Ecology and Evolutionary Biology Princeton University, Princeton, New Jersey, 08544, USA
[3]High Meadows Environmental Institute, Princeton University, Princeton, New Jersey, 08544, USA
[4]Institute for Wearable Convergence Electronics, Department of Electronics and Information Convergence Engineering, Kyung Hee University, Yongin 17104, Republic of Korea
[5]Department of Electronics and Information Convergence Engineering, Kyung Hee University, Yongin 17104, Republic of Korea
[6]Skyhook Bio, LLC, Boston, Massachusetts, 02139, USA
[7]Microsoft Corporation, Redmond, Washington, 98052, USA
[8]Department of Electronic Engineering, Kyung Hee University, Yongin 17104, Republic of Korea
[9]Institute for NanoBioTechnology, Johns Hopkins University, Baltimore, MD, 21218 United States of America
[10]Department of Oncology, Johns Hopkins University, Baltimore, MD, 21218, USA
[11]Sidney Kimmel Comprehensive Cancer Center, Johns Hopkins University, Baltimore, MD, 21218, USA
[12]Hopkins Extreme Materials Institute, Johns Hopkins University, Baltimore, MD, 21218, USA
* Corresponding Author: schur@jhu.edu

The growing use of triaxial particles in microfluidic, microrobotic, and biological systems makes it essential to understand how their rotational dynamics couples with lateral migration in microscale flows. Our experiments in inertial Poiseuille flow reveal that geometric asymmetry in triaxial, multifaceted disks governs their orientation, migration, and rotational period, distinguishing them from classical axisymmetric objects. We identified a Reynolds- and geometry-dependent shift in preferred rotational orientation, arising from the Dzhanibekov effect, with transition modes determined by the particle's principal-axis configuration. We quantified a scalar offset from Jeffery's orbit prediction and introduced a fitting parameter that generalizes the Jeffery equation to include moment-of-inertia effects on rotational dynamics. Finally, we report the diameter of gyration as a predictor of the lateral equilibrium position of inertially focused triaxial particles. Our results link particle asymmetry to migration and rotation in flow, expanding our understanding of particle dynamics.

Microscale triaxial particles - objects with three unequal principal moments of inertia - have shifted from theoretical curiosities to practical micro-components due to advances in lithographic and 3D-printing techniques that enable fabrication of fully three-axis-asymmetric shapes at sub-micron resolution [1–3]. These particles are increasingly used in microfluidic [4], micro-robotic [5], biological (e.g. diatom [6]), and colloidal systems [7], where viscous and inertial forces govern their translational and rotational behaviors, yet the hydrodynamic principles underlying their dynamics remain largely unexplored.

Even in the absence of flow, triaxial asymmetry gives rise to complex, often unstable rotation [8–10], exemplified by the Dzhanibekov effect, where spinning about the intermediate inertia axis induces spontaneous flipping [10–12]. While well understood in classical rigid-body mechanics, such instabilities remain poorly characterized in viscous or inertial flows, particularly at the microscale where particle shape, confinement, and flow field are tightly coupled [13–15]. Consequently, prior studies have been constrained to axisymmetric particles, which serve as simplified models for the more complex dynamics of truly triaxial shapes. Accordingly, the motion of axisymmetric ellipsoids and cylinders has been extensively investigated in microflows [13,16–19]. Jeffery's classical analysis demonstrated that the periodic rotation of ellipsoids in Stokes Couette flow could be modeled as a function of the shear rate and aspect ratio [20], laying the foundation for understanding non-spherical particle motion in fluid [20]. Subsequent studies extended Jeffery's framework to Poiseuille inertial flow [13,16–19] and for cylindrical particles [7,21,22]. At the opposite limit, under creeping flow conditions, prior work has shown that triaxiality (slight deviations from symmetry) can induce chaotic, quasi-periodic, or double-periodic rotation in triaxial ellipsoids and cylinders [13,17,27–29]. However, beyond qualitative dependence on the degree of triaxiality and initial orientation, these works do not provide a generalizable framework for predicting the rotational orientation of more complex triaxial particles in flow [13,17,27–29].

Despite extensive understanding of axisymmetric particle rotation and inertial focusing, the coupling and focusing mechanisms of triaxial particle remain unclear [13,15,23–26]. In a rare example, Hur *et al.* [26] found that most triaxial doublets and cylinders in inertial Poiseuille flow focused at lateral equilibrium positions predicted by their rotational diameter, $D_{max}$. Tohme *et al.* [13] later showed that this observation agreed with three other independent studies on ellipsoidal particles [16,30,31]. The only deviations involve soft, deformable biological particles, which experience additional lift forces due to deformability in Poiseuille flow [31–35], and the "h-shaped" multi-faceted disk (h-disk) [26]. This finding suggests that the lateral focusing of h-disks and other non-deformable triaxial particles may be influenced by subtle geometric features beyond rotational diameter and aspect ratio.

Here, we investigated the rotational dynamics and lateral equilibrium positions of six custom-fabricated, inertially focused polymeric multifaceted disks (MFDs)—triaxial particles with a range of triaxialities and symmetry classes. First, we show that particle triaxiality profoundly affects the preferred rotational orientation. Secondly, we introduce a geometry-corrected formulation of Jeffery's equation that describe the rotational period of MFDs. Finally, we demonstrate that the diameter of gyration serves as a predictive metric for the lateral equilibrium position of MFDs.

Triaxial MFDs were fabricated using a stop-flow lithographic method previously described [26] to create six distinct shapes with cross-sections resembling the letters U, H, V, S, J, and L (FIG. 1). The rotational diameter ($D_{max}$) and aspect ratio ($AR$) were fixed at approximately 0.6 to decouple the effect of shape from those of size [16] and aspect ratio [17], matching the parameters used for the h-shaped MFD reported by Hur *et al.* [26].

In our experimental setup, MFDs were suspended in 200 proof ethanol (Sigma-Aldrich) and injected into a tall rectangular PDMS microchannels ($W \times H = 115 \times 143\mu m^2$, $H/W \approx 1.5$) using syringe pump (Harvard Apparatus PHD Ultra) at nominal flow rates of 370, 916, 1100, and 1466μL/min, corresponding to channel Reynolds numbers (defined in supplementary section 2 [37]) ≈ 32, 78, 94, and 125. By harnessing the slight compliance of PDMS [36], each setpoint naturally produces a gentle ramp in flow rate so that over the course of empirical testing particles experienced a continuous channel Reynolds number sweep from 15 up to 240 (FIG S1 [37]). This smooth variation allowed us to characterize MFD behavior across the full inertial regime without discrete jumps between conditions. We collected videos of particles 4.5cm downstream of the inlet of our device using our high-speed video set up [26]. We utilized a custom MATLAB code which enables accurate detection of particle inertial focusing behavior and geometric features. (Supplementary section 4-5 [37] ).

*Contact author: schur@jhu.edu

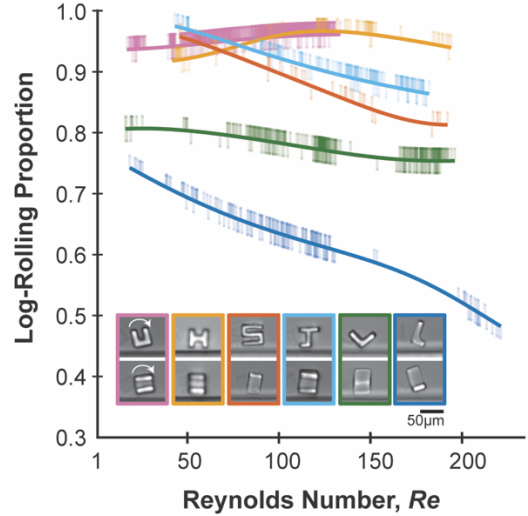

FIG. 1 Triaxiality-driven changes in the rotational orientation of MFDs. Proportion of particles in log-rolling orientation as a function of Reynolds number. Stem plots from the trend line indicate log-rolling (above) and tumbling (below) orientations. Inset images show high-speed microscopic snapshots of flowing MFDs in log-rolling (top) and tumbling (bottom) orientations.

Inertially focused triaxial MFDs exhibited two predominant Jeffery-like orientations: log-rolling (rotation about the extrusion axis) and tumbling (rotation orthogonal to the extrusion axis), together comprising over 85% of observed rotations. Since particle orientation of triaxial particles is known to be sensitive to minor defects [13,15,23–26], we excluded inclined particles (<15%) from further quantitative analysis, as this orientation likely arose not from the intended MFD shape but from defects – both visible (observed in 35% of inclined particles (FIG S6, Table S2 [37])) and undetected ones possibly hidden in the field of view.

The presence of only two orientations of rotation in MFDs contrasts with axisymmetric ellipsoids and cylinders reported to exhibit five or more orientations (log-rolling, tumbling, kayaking, inclined-rolling, and steady-state) across Stokes and inertial regimes in shear or pressure-driven flow [13,17,27–29]. Notably, whereas triaxial ellipsoids [13,15,23–25], faceted triaxial prolate cylinders [23], and other triaxial particles [10,19,23,39] in Stokes-regime Couette flow often exhibit chaotic or double-periodic rotations depending on asymmetry and initial orientation, MFDs maintained only the two Jeffery orbits typical of oblate and prolate ellipsoids [13,38]. This contrast suggests that the inertia effects governing the motion of MFDs suppress chaotic dynamics and reinforce the dominant Jeffery orbits.

We quantified the rotational preference between the two Jeffery-like orbits using Gaussian kernel density estimation (Supplementary Section 8 [37]) as a function of Reynolds number (FIG. 1). At low Reynolds number, over 75% of MFDs—regardless of shape—adopted a log-rolling orientation, consistent with prior observations for disk-like cylinders and ellipsoids [13,17,29]. As Reynolds number increases, MFDs extruded along their maximum principal axis (H, U) remain in the log-rolling regime, whereas those extruded along their intermediate principal axis (L, J, V) undergo a clear transition from log-rolling to tumbling. The observed transitions are consistent with the Dzhanibekov instability [10–12], wherein rotation about the intermediate axis becomes unstable, inducing spontaneous axis flipping. The S-shaped MFD, although extruded along its maximum axis, appears to exhibit a similar transition; however, the limited number of observations (n < 30) prevents definitive conclusions (<75% confidence, see Supplementary Section 8 [37]).

Building on these observations, we related the rotational orientation of MFDs to their geometric characteristics and moment of inertia ratios. Similar to other axisymmetric non-spherical particles [13], the log-rolling axis corresponds to the principal axis with the smallest protrusion distance, $D_{ext}/W$, (FIG A 2c). The tumbling axis, in contrast, corresponds to the principal axis most distinct from the log-rolling principal axis (the largest principal momentum transition, $\xi$, defined in Appendix A, FIG A 2f). These relationships define a geometric framework for identifying which principal axis serves as the log-rolling and tumbling axis of any MFD.

We next examine geometric measures that correlate with the prevalence and transition dynamics between log-rolling and tumbling orientations. The baseline proportion of MFDs in log-rolling (e.g. absolute proportion at Re ~100, FIG. 1) was inversely correlated with particle anisotropy, defined as the ratio of the maximum to minimum principal moments of inertia ($\alpha = \frac{I_{max}}{I_{min}}$; FIG A 2). In contrast, the rate of transition from log-rolling to tumbling with increasing $Re$ (i.e. the slope in FIG. 1) was governed by triaxiality ($\tau$), defined as the Euclidean distance to isotropy (FIG A 2). Together, these results indicate that greater deviation from symmetry – whether from isotropy or from a symmetric-top configuration (i.e. two of the three principal axes are equal) – promote tumbling at higher $Re$. Although further work is needed to clarify the coupling between geometry and rotational inertia, these results establish a predictive framework linking particle triaxiality and anisotropy to orientations stability and transitions across Reynolds numbers, even for complex or asymmetric particle shapes.

To extend this geometric framework to rotational dynamics, we next adapt Jeffery's classical period formula, $T = \left(AR + \frac{1}{AR}\right)\frac{2\pi}{\dot{\gamma}}$ [20], to predict the rotational period for triaxial MFDs. Here, $AR = d/D_{max}$ is the aspect ratio of MFDs where $d$ is the extruded depth and $D_{max}$ is the rotational diameter in the log-rolling orientation. The local shear rate is defined as $\dot{\gamma} = 2v/Y$, where $v$ is the particle velocity and $Y$ is the lateral equilibrium position from the nearest wall, both obtained from high-speed video analysis.

While Jeffery's equation was initially derived for axis-symmetric ellipsoids, prior work has introduced corrections for cylindrical particles by multiplying the aspect ratio ($AR$) by a factor, $c$, typically $c$=0.7 for prolate rods [7,21,22] and 1.6<$c$<2.3 for oblate disks [40], to account for geometric drag differences [7].

Despite exhibiting cylindrical, disk-like inertial-focusing behavior (Appendix B, FIG A 3), triaxial MFDs show shape- and orientation-dependent deviations from Jeffery's predictions (FIG S13 [37]) that cannot be corrected by a single linear $AR$ adjustment as done for cylindrical particles [7,21,22,40]. To account for the geometry-dependent rotational inertia and complex rotational dynamics of MFDs, we formulated a geometry-dependent Jeffery equation that incorporates the ratio of reference to effective rotational inertia, $T = C_m \frac{I_o}{I_{rot}}\left(AR + \frac{1}{AR}\right)\left(\frac{2\pi}{\dot{\gamma}}\right)$ where $C_m = 2$ is an empirically determined scaling factor, $I_{rot}$ is the moment of inertia of an MFD about the active rotation axis, and $I_o$ is the reference moment of inertia defined by its bounding rectangular prism (FIG. 2 inset schematic). $I_{rot}$ was calculated by reconstructing three-dimensional particle geometry from facet lengths extracted from integrated images of MFDs observed in both log-rolling and tumbling states (Supplementary Section 6 [37]). The reference inertia, $I_o$, was computed as $I_o = \frac{1}{12} M (h^2 + w^2) = \frac{\rho}{12} hwd(h^2 + w^2)$, where $h$, $w$, and $d$ are the height, width, and depth of the bounding rectangular prism following the edges of each MFD. The dimensionless ratio, $C_m \frac{I_o}{I_{rot}}$, therefore, quantifies how an MFD's facets and cavities modify its rotational inertia—effectively measuring how these geometric features increase or decrease the particles' resistance to rotation relative to a solid rectangular prism. For simplicity, the tumbling axis of the reference rectangular prism ($I_o$) was assumed to align with the nearest particle edge rather than an offset axis. This approximation was necessary for shapes such as the V and J, where the actual rotation axis can be offset from any individual edge and may vary between particle instances. For example, a V-shaped MFD may tumble about an axis approximately 45° offset from one of its log-rolling edges (Fig. 1, inset), or with little to no offset, resembling the tumbling L-shaped particle shown in the side-view inset of FIG. 2 (Supplementary video [37]). Given this variability—and the limited three-dimensional information available from our image analysis pipeline—we adopted a bounding-box-based estimate of the rotation axis for calculating $I_o$.

*Contact author: schur@jhu.edu

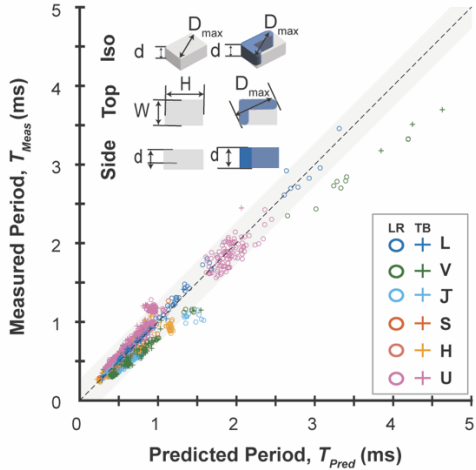

FIG. 2 Inertia-corrected Jeffery's orbit period prediction for triaxial multifaceted disks. Measured vs. predicted rotational period ($T_{meas}$ vs. $T_{pred}$) showing strong agreement within the 95% confidence interval (shaded region). Inset schematics illustrate geometric parameters used for modification, and marker symbols denote different MFD geometries and orientations (legend).

Compared to the classical Jeffery equation, the geometry-corrected form significantly improved agreement with experimental rotational periods, achieving $R^2 > 0.8$ (average $R^2_{ave} = 0.89$) for all cases except the S-tumbling and J-shaped MFDs (FIG. 2, Table S10 [37]). The consistency across multiple shapes and orientations confirms that incorporating the geometry-dependent inertia ratio ($C_m \frac{I_o}{I_{rot}}$) substantially improves predictive accuracy over the classical Jeffery model.

The S-tumbling and J-shaped MFDs showed lower fits ($R^2 = 0.45$ and $R^2 = 0.40$-$0.77$, respectively) due to fabrication inconsistencies and limited sample sizes (Table S1 [37]). Additional discussion of these effects is provided in Supplementary Section 11 [37]. Because direct 3D imaging was not utilized, moments of inertia were estimated from reconstructions of average particle geometries. These approximations–along with assumptions regarding bounding-box alignment in tumbling states—introduce a consistent multiplicative bias across all shapes, as each is scaled by the same reconstructed factor ($C_m \frac{I_o}{I_{rot}}$). This systematic bias produces a linear offset in the fitted relationship and likely explains why the V-shaped MFD, despite achieving strong fits ($R^2 = 0.87$ and $0.86$ for log-rolling and tumbling, respectively), appears slightly misaligned in FIG. 2 due to its edge-offset tumbling axis.

*Contact author: schur@jhu.edu

Despite these limitations, the proposed model successfully captured the influence of complex mass distribution on MFD rotation. Whereas Jeffery's original framework predicted particle rotation solely from hydrodynamic stresses on the surface of the particle [20], our results reveal that the moment of inertia plays a critical role in determining the rotational period of triaxial MFDs. Although this geometry-weighted approach relies on high-fidelity 3D shape reconstructions, it offers a generalizable tool for engineering particle geometries that achieve desired rotational behavior in inertial microflows [7]. Yet, the connection between rotational dynamics and lateral migrations in inertial flows has not been fully established.

Recent studies in inertial focusing have indicated that particle rotation can generate a rotation-induced lateral lift force, which shifts the equilibrium position of spherical particles [41]. Motivated by this coupling between rotation and lateral migration, we tested whether the established linear model, $X_{eq}=0.54 (D_{max}/W)+0.34$–initially reported by Hur *et al.* [26] and subsequently validated by Tohme *et al.* [13] and other independent studies [16,30,31]–accurately predicts the lateral focusing positions of MFDs. MFDs did not follow this model, consistently focused closer to the channel wall than predicted (FIG. 3a) and exhibiting a significantly weaker dependence on $D_{max}/W$ (slope = $0.18 \pm 0.07$, FIG A 3d). To further elucidate the role of particle cross-sectional shape in lateral focusing, we fixed $D_{max}/W=0.634$ (value for h-disk [26]) and plotted $X_{eq}$ as a function of cross-sectional elongation ($E$). Elongation was selected as an additional geometric metric because prior studies have shown that for axisymmetric prolate ellipsoids, the equilibrium focusing position shifts toward the channel center as elongation increases [16]. Here, elongation ($E$) is defined as the ratio of the major and minor axes of an ellipsoid fit to the extrema of each MFDs (inset, FIG. 3 b). For faceted particles, this measure captures in-plane stretching while remaining independent of internal cavities or surface faceting. For tumbling ellipsoids, $E$ is equivalent to the aspect ratio used in previous studies [13].

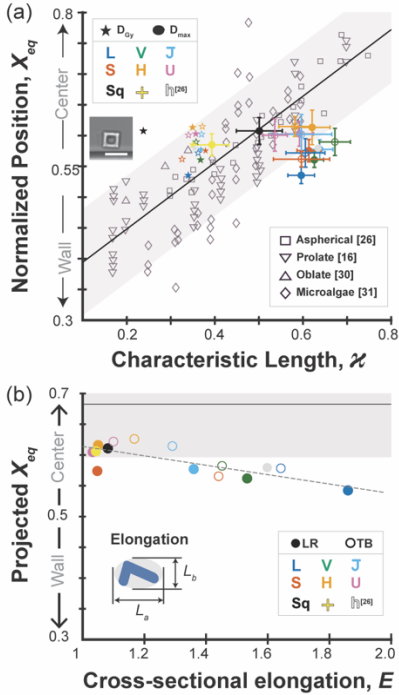

FIG. 3 Lateral equilibrium position of MFDs. (a) Normalized equilibrium position ($X_{eq}$) as a function of characteristic length ($\kappa$). Colored markers represent experimental data from this study: circles correspond to the rotational diameter ($D_{max}/W$), and star-shaped markers correspond to the diameter of gyration ($D_{gy}$). Open gray markers indicate previously reported data for various axisymmetric particles. The solid black line ($X_{eq} = 0.54(D_{max}/W) + 0.34$) was originally reported by Hur *et al.* [26]; the shaded region indicates the confidence zone, defined as half the radius of the smallest particle, following Tohme *et al.* [13]. Inset shows a representative image of the square-ring (Sq); scale bar, 50μm. (b) Projected equilibrium position ($X_{eq}$) at constant $D_{max}/W = 0.634$ as a function of cross-sectional elongation ($E$). Dashed line indicates linear fit to data. Solid and open markers denote log-rolling and tumbling orientations, respectively.

Unlike axisymmetric prolate ellipsoids, our MFDs focused progressively closer to the channel wall with increasing $E$. This divergence likely arises because previous studies varied both $E$ and $D_{max}$ simultaneously, whereas our approach isolates the contribution of shape by holding $D_{max}$ constant. Notably, MFDs with $E>1.4$ (including h-disk at $E \approx 1.6$. [26]) fall outside the confidence zone (defined by Tohme et al. [13]) of the established linear scaling (shaded region in FIG. 3a), demonstrating that $D_{max}$ alone is insufficient to predict the focusing behavior of complex particle shapes.

To capture both shape complexity and rotational resistance, we evaluated the diameter of gyration, defined as $D_{Gy} = 2\sqrt{\frac{I_{rot}}{M}}$, where $I_{rot}$ is the rotational moment of inertia and $M$ is particle mass. While $D_{max}$ and $AR$ were held approximately constant as 0.6, the MFD cross-sections span a wide range of elongation ($E$), extrusion-axis orientation, symmetry classes, and moment-of-inertia ratios. For validation, we evaluated two additional shapes—a plus (+) and a square-ring (Sq)—with smaller $D_{max}/W$ (~0.4 and 0.5) and thinner profiles ($AR \approx 0.25$) (FIG. S14b). These shapes, which differ in size and aspect ratio, were used to assess whether the diameter of gyration ($D_{Gy}$) serves as a universal predictor of lateral equilibrium position across MFDs with varying shape, symmetry, and size. When plotting $X_{eq}$ against $D_{gy}/W$ (star-shaped markers in FIG. 3a), nearly all MFD data—including both log-rolling and tumbling orientations—collapsed onto the previously reported linear trend. The square-ring was the sole outlier, likely because its large central void—absence in all other MFDs—modifies local lift forces [42,43].

The stronger performance of $D_{gy}$ across triaxial MFDs highlights the importance of geometry-dependent rotational dynamics in determining lateral equilibrium position. $D_{gy}$ extends the relationship between geometry and inertial focusing; this result is consistent with the complex internal asymmetries that govern the rotational dynamics of MFDs, as shown in FIG. 1 and FIG. 2. This insight reinforces that moment of inertia governs not only how particles rotate, but also where they migrate in flow.

In summary, our findings reveal that triaxiality and geometry-dependent rotational dynamics strongly influence the inertial focusing behavior of polymeric multifaceted disks. We demonstrate that Jeffery's theory–despite its long history and successful extensions to axisymmetric particles theory [7,13,14,20,22]–is insufficient for particles lacking rotational symmetry. We introduce a generalizable, physically grounded framework that integrates inertia-based corrections and geometry-weighted moments of inertia to predict and design complex particle dynamics.

While our experiments focused on shape-locked polymeric disks in dilute Newtonian flow, a small fraction of particles (4-15%) exhibited non-Jeffery "inclined" orbits whose origin – whether from fabrication imperfections or previously unreported orientation - remains to be resolved. Extending the present inertia-weighted framework to soft or concentrated suspensions will require dedicated studies of deformation- and interaction-driven forces. Future efforts should explore collective behaviors of triaxial particles in concentrated or viscoelastic suspensions – where inter-particle interactions and non-Newtonian stresses may yield emergent alignment or novel rheology and investigate how particle softness or surface texture modify inertial focusing and rotation. Together, these directions will expand the predictive and design capabilities of our geometry-informed framework for triaxial particle dynamics [44,45].

*Contact author: schur@jhu.edu

**Acknowledgements:**
This research was supported by the Rowland Fellowship from Harvard University and by the National Science Foundation under Grant No. 1804004. We would like to acknowledge editing support by Anne N. Connor, provided by the Office of the Vice Provost for Research at Johns Hopkins University. ChatGPT, a language model developed by OpenAI, was utilized to improve clarity of the manuscript's wording and to generate pseudocode in select analyses.

**Data Availability:**
The data that support the findings of this study are available from the corresponding author upon reasonable request.

*Contact author: schur@jhu.edu

*Contact author: schur@jhu.edu

# Appendix A: Triaxial MFD Library

We fabricated polymeric triaxial multi-faceted disks (MFDs) with six cross-sectional shapes, L, V, J, S, H, and U – each engineered to share the same rotational diameter, $D_{max}$, and the Jeffery aspect ratio, $AR=d/D_{max}$, where $d$ is the extrusion thickness (FIG A 1b). Controlling both $D_{max}$ and $AR$ lets us isolate the effect of triaxial shape (rather than size or aspect ratio) on inertial focusing, in contrast to prior studies that varied these parameters simultaneously [13,16,17]. All MFDs were designed with $AR \approx 0.6$ (disk-like morphology) and $D_{max} \approx 0.6$ (FIG A 1), matching the $D_{max}$ (0.635) of h-disk studied by Hur *et al.* [26], whose $AR$ was unspecified. Previous studies identified $D_{max}$ as a key predictor of the lateral equilibrium position of inertially focused aspherical particles [13,16,26,30,31], yet the coupled variation of shape and size ($AR$ and $D_{max}$) in those works hindered clear isolation of their individual effects [16]. This distinction is important because oblate (disk-like, *low-AR*) and prolate (rod-like, *high-AR*) particles exhibit distinct rotational and focusing dynamics [17].

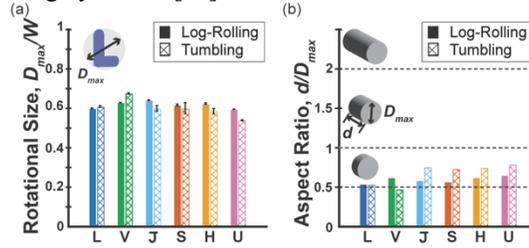

FIG A 1 Size- and Aspect- Ratio-Controlled MFDs. To isolate the effect of shape on inertial focusing, MFDs were designed to have (a) similar rotational diameter across shapes ($D_{max} \approx 0.6$) and (b) disk-like aspect ratio ($AR<1$).

Our six MFD shapes span a range of triaxialities and symmetry classes (FIG A 2): V, U, and H have mirror symmetry; S, H, and U possess discrete axial symmetry; and L and J are fully asymmetric. Particles were oriented as if their cross-section were extruded along either the intermediate (L, V, J; $I_{int}$) or maximum (U, H, and S; $I_{max}$) principal axis. To better understand how particle geometry and moment of inertia influence the preferred rotational orientation, we quantified the protrusion distance ($D_{ext}$), anisotropy ($\alpha$), triaxiality ($\tau$), and principal moment transition ($\xi$). $D_{ext}$ indicates the mass range along the direction of the principal axis (FIG A 2 c). Anisotropy ($\alpha$) is the ratio of maximum to minimum principal moments of inertia ($\alpha = I_{max}/I_{min}$) effectively comparing the degree to which partilces deviate from a spherical particle not accounting for possibility of two of three principal axis being equal. Triaxiality ($\tau$) is defined as the Euclidean distance from isotropy $\tau = \sqrt{\left(\frac{I_{min}-I_{avg}}{I_{avg}}\right)^2 + \left(\frac{I_{int}-I_{avg}}{I_{avg}}\right)^2 + \left(\frac{I_{max}-I_{avg}}{I_{avg}}\right)^2}$. Principle moment transition ($\xi$) is defined as $\xi = \frac{I_{tumb}-I_{Log}}{I_{Third}-I_{Tumb}}$, where $I_{Log}$, $I_{Tumb}$, and $I_{Third}$ are the principle moment of inertia about the log-rolling, tumbling, and remaining orthogonal axes, respectively.

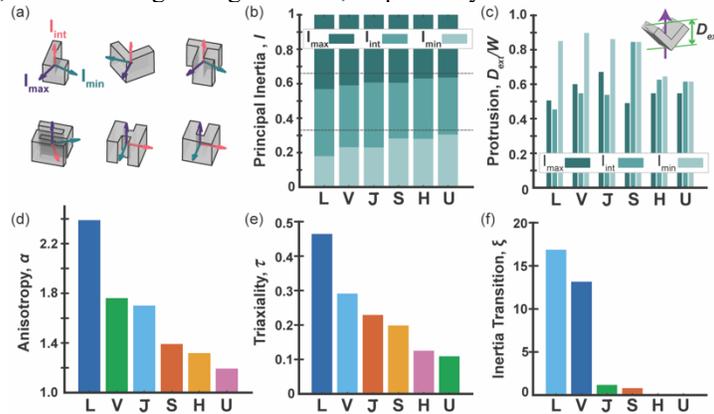

FIG A 2 Asymmetry Characterization of MFDs. (a) Three principal moments of inertia for each MFD shape. (b) Proportion of MFD principal axis. (c) Protrusion distance for each MFD geometry. (d) Anisotropy ($\alpha$), defined as the ratio of the largest to smallest principal moment of inertia. (e) Triaxiality ($\tau$), defined as the Euclidian distance from isotropy. (f) Principal moment

*Contact author: schur@jhu.edu

transition ($\xi$), defined as the change between the principal moment of inertia corresponding to the initial orientation axis (e.g., log-rolling) and that corresponding to the final orientation at high Reynolds number normalized by the difference between the initial and third principal moments of inertial.

## Appendix B: Stable Equilibrium and Lateral Confinement of MFDs in Inertial Flow

All MFD shapes migrated to two lateral equilibrium positions at the channel mid-height (H/2), translating with a velocity coefficient of variation (CV) below 0.1 over $15 < Re < 240$ (FIG A 3a), consistent with the inertial-focusing behavior previously reported for spheres and asymmetric particles in tall rectangular channels [13,16,17,26,46–49]. Uniform image contrast, characterized by a bright center with dark edges in each particles [41], further confirmed that particles were confined to a single focal plane, demonstrating strong inertial focusing-induced vertical confinement.

Inertially focused MFDs exhibited lateral oscillations in their equilibrium position ($X_{eq}$) with amplitudes up to $0.15D_{max}$, corresponding to $\Delta X_{eq} \approx 0.2$ (FIG A 3b). These values exceed the $\sim 0.0125D_{max}$ oscillations reported for rod-like cylinders ($AR = 2$, $Re = 100$) [17], likely due to the uneven mass distribution and triaxial geometry of MFDs, which promote coupled rotational-translational dynamics. Despite these oscillations, MFDs maintained stable mean $X_{eq}$ even at Reynolds numbers as low as 16—well below the $Re > 30$ threshold typically required for non-spherical cylinders to focus [26]. Over the range $15 < Re < 240$, the mean lateral variation ($\overline{\Delta X_{eq}}$) remained below 0.05 (FIG A 3c), consistent with the weak $Re$-dependence observed by Su *et al*. [17] for thin cylindrical disks ($AR = 0.26$, $\overline{\Delta X_{eq}} < 0.01$ over $50 < Re < 200$). Together, these results indicate that, despite their geometric complexity, MFDs behave similarly to disk-like particles in their equilibrium positioning under inertial flow.

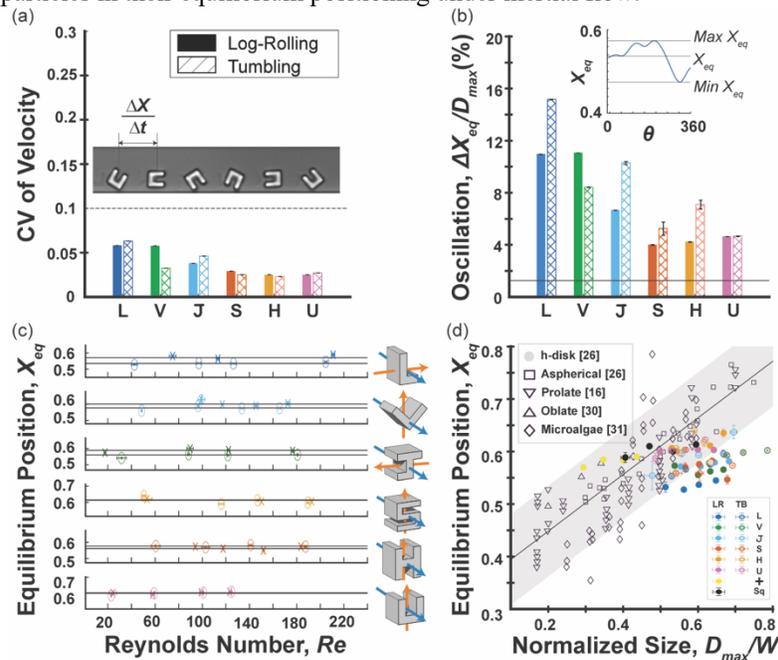

FIG A 3 (a) Inertial focusing behavior of MFDs. (a) Inertial focusing was assessed from high-speed video analysis, where particles were considered focused when the coefficient of variation (CV) of their single-particle velocities in the field of view (FOV) was less than 10%. (b) Fluctuations in lateral equilibrium position ($X_{eq}$) during a single rotational cycle for individual MFDs. The dashed line indicates a 1.25% fluctuation threshold previously reported for ellipsoidal particles [13]. (c) Equilibrium lateral position ($X_{eq}$) as a function of Reynolds number, shown separately for each MFD shape. Data are grouped into four bins with standard error bars. (d) Normalized equilibrium position ($X_{eq}$) as a function of the normalized rotational diameter ($D_{max}/W$) for MFDs compared to prior studies, as summarized by Tohme *et al*. 2021 [13]. Inset: plus (+) and square-ring (Sq) shapes from this study represent particles with a lower aspect ratios ($AR \approx 0.25$) and smaller rotational diameters ($D_{max}/W \sim 0.4$ and 0.5).

*Contact author: schur@jhu.edu

# Supplementary - Triaxial Asymmetry Driven Rotational Dynamics and Lateral Equilibrium Position In Inertial Flow

## I. SUPPLEMENTARY MATERIALS

### 1. Particle Synthesis

Multifaceted Disks (MFDs) with cross-sectional shapes resembling letters (L, V, J, S, H, U, +, and square-ring) were fabricated using an optofluidic maskless lithography setup combined with stop-flow lithography as previously described [1,2]. Briefly, in our experiments we introduced a photo-curable oligomer mixture, Poly (ethylene glycol) diacrylate (PEGDA, MW = 700, Sigma-Aldrich) with 10% v/v of photo initiator (2-Hydroxy-2-methylpropiophenone, Sigma-Aldrich) into a microfluidic channel whose height was matched to the designed particle depth (*d*) to be fabricated. Ultraviolet light patterns reflecting desired particle shapes were modulated by a digital micromirror device (Texas Instruments) and projected through the microfluidic channel filled with the photo-curable oligomer mixture.

### 2. Inertial Focusing of Triaxial MFDs

Straight rectangular microchannels with a high aspect ratio ($W \times H = 115 \times 143 \mu m$, aspect ratio, $H/W \approx 1.5$) were fabricated using standard photolithography techniques, as described previously [1]. The MFDs were suspended in 200 proof-ethanol (Sigma-Aldrich) to prevent particle aggregation caused by residual uncured oligomer and then sonicated using an ultrasonic probe to reduce clumping before experimentation. MFDs were injected into the microchannel using syringe pumps at nominal flow rates of 370, 916, 1100, and 1466µL/min, corresponding to channel Reynolds numbers $Re = \frac{\rho v W}{\mu} \approx$ 32, 78, 94, and 125 (FIG S 1). Here, $\rho$ is the density of fluid (ethanol: 785.04g/cm³ [4]), $\mu$ is its dynamic viscosity of ethanol (1.074mPa·s [4]), $W$ is the microfluidic channel width, and $v$ is the particle velocity obtained from image analysis. Rather than estimating the channel Reynolds number, $Re_c = \frac{\rho U_m W}{\mu}$, from the syringe pump's bulk flow rate ($U_m$), we computed $Re$ using particle's translational velocity, $v$, to (1) directly assess how local velocity influences particle behavior and (2) capture the local flow conditions experienced by individual particles (FIG S 1). Although the particle Reynolds number, $Re_p = \frac{\rho U_m a^2}{\mu W} = Re_c \times \left(\frac{a}{W}\right)^2$, is the proper metric for local shear, its quadratic sensitivity to slight diameter variations – whether from manufacturing tolerances or measurement/detection error - can yield vastly different $Re_p$ for particles with similar velocity, thereby conflating dimensional variability with true hydrodynamic effects. Dimensional variability was independently assessed (see SI section 5, FIG S 7, Table S 3-6) and found to exert only a minimal impact relative to the hydrodynamic effects under study. All particle motion was recorded at 4.5 cm downstream from inlet – well beyond the critical channel length $L_f$ (1.7mm for the lowest tested flow rate) – to ensure that particles had reached their equilibrium positions before imaging with the high-speed imaging system described in Supplementary section 3. The critical length $L_f$ was calculated as $L_f = \frac{\pi \mu W^2}{\rho v D_{max}^2 f_L}$ [3] where $\mu$ is the dynamic viscosity of ethanol 1.074mPas [4], $W$ is the microfluidic channel width, $\rho$ is the density of ethanol 785.04g/cm³ [4], $v$ is the particle velocity obtained from image analysis, $D_{max}$ is the rotational diameter of our particle, and $f_L$ is the average lift coefficient set to be 0.035 [5].

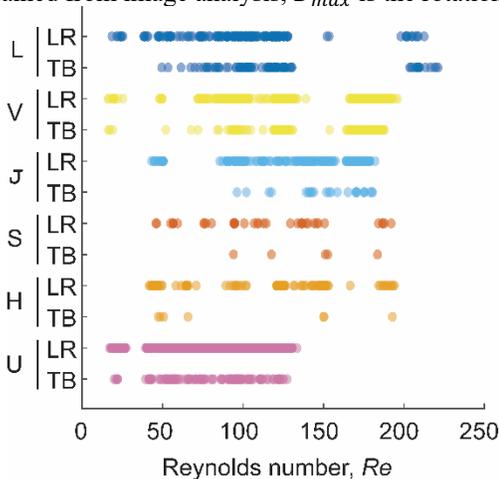

**FIG S 1:** Observed orientation states of MFDs as a function of Reynolds number ($15 < Re < 240$). Each point represents an individual particle observation, with colors corresponding to particle shape. LR and TB denote log-rolling and tumbling orientations, respectively.

### 3. Optical Architecture for High-Speed Microscopic Analysis

We constructed a high-speed imaging setup with ultra-bright illumination to capture kinematic information from relatively transparent polymeric microparticles moving at over 1m/s – equivalent to more than 16,000 body lengths per second [1]. The apparatus can sequentially collect high-contrast images without motion blurs using a high-speed camera (Phantom® v1210, Vision Research, Inc.). We engineered the illumination apparatus such that the CMOS sensor of the high-speed camera with a 1µs aperture time could capture sharp, crisp images of the flowing objects through a 40X objective (N.A.= 0.6 and FOV= 22mm) on the inverted Epi-fluorescent microscope (Eclipse Ti, Nikon). The 150 W Xenon light source (Oriel® Instrument) generates an intense light ray with an irradiation wavelength ranging from 200 to 2400nm. We installed the water filter and the ultraviolet bandpass filter in the light path of the source for only visible wavelengths to transverse the region of interest and be collected by the CMOS sensor, while minimizing the transmission of thermal energy to the sample. We collimated the intense source ray by utilizing a minimal number of lenses and optical apertures to maximize the illumination intensity. A significant improvement in image quality was evident when the CMOS camera captured flowing droplets at >30,000 frames per second (FIG S 2).

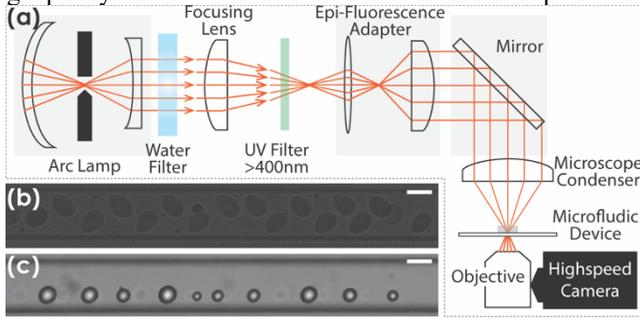

**FIG S 2:** (a) Optical setup designed to enhance image contrast during high-speed imaging of microscale objects moving at extreme velocities. Representative high-speed microscopic images of flowing microscale objects (b) before and (c) after contrast enhancement. All the scale bar: 50μm.

### 4. Image Analysis of Recorded Videos

We developed custom MATLAB code to semi-automatically extract the geometric (FIG S 3) and kinematic features (FIG S 4) from high-speed videos (15,000 to 40,000 frames per second, ~100,000 frames per video), enabling efficient analysis of these large datasets. Briefly, the MATLAB code employs a background subtraction method to isolate frames containing MFDs. On average, fewer than 5% of the frames contain MFDs, as the particle concentration was intentionally kept low to minimize particle-particle interactions [6]. Frames containing detected particles were cropped to isolate individual MFDs. Particle outlines were then determined using an algorithm adapted from MathWorks [7]. The cropped images were analyzed to determine particle rotational orientations (log-rolling, tumbling, or inclined FIG S 3), feature sizes (FIG S 4), position (center of area FIG S 4), and angle of rotation (FIG S 4). To quantify the angular displacement of rotating particles between two consecutive frames, we rotated the particle's reference image, $P_{Ref}$, by a candidate angle, $\theta$, and compute the cost function $C(\theta) = \frac{1}{2}\sum(P_{Ref}(x,y,\theta) - P_{Obs})^2$ where $P_{Obs}$ is the current frame. The estimated rotation $\Delta\theta$ is the value of $\theta$ minimizing $C(\theta)$. The cropped images were manually inspected, and any particle meeting one or more of the following five criteria was excluded from further analysis: (i) presence of defects in the particle (FIG S 5a), (ii) proximity to another particle (< 90 pixels or < 2X particle diameters FIG S 5b), to prevent particle-particle interactions [6], (iii) interference from background debris, (FIG S 5c) (iv) inclined rotational orientation (FIG S 5d), or (v) inconsistent tracking across consecutive frames. (e.g., abrupt jumps in detected positions or orientations, such as when a track switches to a different particle in FOV). This manual curation ensured that only well-formed, isolated particles with reliable tracking and Jeffery-like rotational behavior were analyzed. Also, automatically acquired shape-related geometric features were visually inspected for accurate detection and manually reanalyzed for cases of poor detection. Several inclined particles exhibited minor defects and were noted for further discussion in the main text (FIG S 6, Table S 2).

Our experiments were conducted in a tall, rectangular microchannel using a bottom-up imaging configuration which enabled the visualization of particles at both the top and bottom edges of the field of view. These positions correspond to the two lateral inertial focusing locations in the vertical center plane of the channel. Due to the vertical symmetry about the channel centerline, particles at opposing equilibrium positions exhibited mirrored rotational behavior, rotating in opposite directions despite the unidirectional flow. While this mirroring effect does not influence the rotational analysis of particles with planar symmetry (such as spheres, oblate or prolate ellipsoids, or symmetric MFDs), it must be considered for asymmetric MFDs. In such cases, the data should be reflected prior to analysis to correct for this mirroring. This step is especially critical for particles with L and J shapes may also across the channel's centerline. The mirrored cross-sectional shape can lead to mirrored behavior in terms of rotational velocity, where the highest velocity may occur at either the beginning or the end, depending on how the particle's orientation aligns with the shear direction (further discussed in SI section 9).

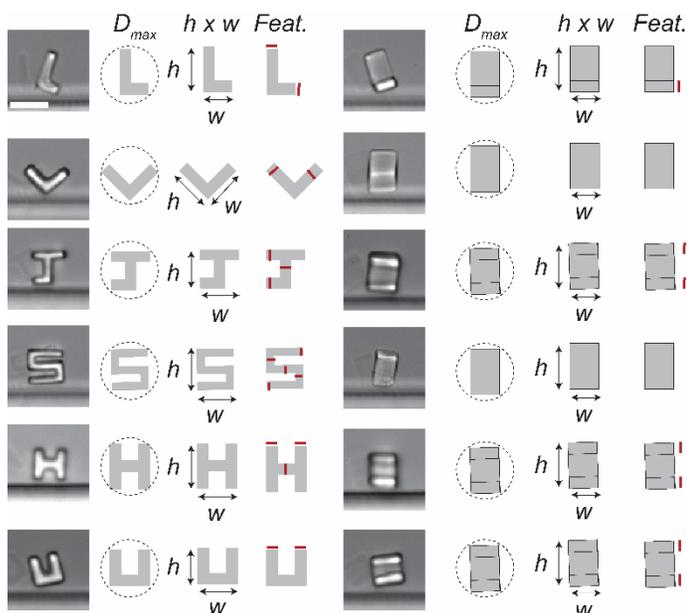

**FIG S 3:** Particle geometries and corresponding dimensions of MFDs observed in the field of view during log-rolling (left) and tumbling (right). rotational diameter ($D_{max}$), height ($h$), width ($w$), and facet locations (denoted 'Feat.' and marked in red) are indicated for each shape. Scale bar is 50μm.

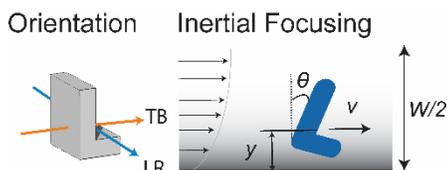

**FIG S 4:** Parameters extracted from the image-processing pipeline for inertial focusing analysis, including particle orientation (LR or TB), angle displacement ($\theta$), distance from the channel wall ($y$), particle velocity ($v$), and channel width ($W$).

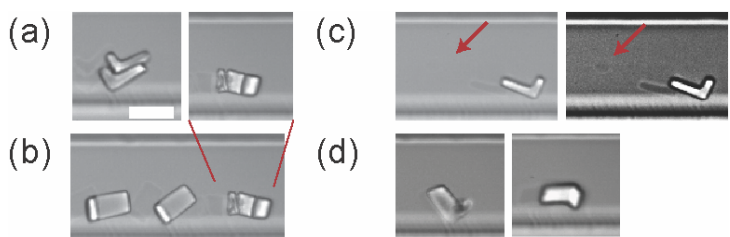

**FIG S 5:** Exclusion criteria for image analysis. (a) Particle defects. (b) Proximity to other particles. (c) Background debris (left: original image; right: contrast-adjusted image; red arrows indicate debris). (d) Inclined or out-of-plane particle orientations. Scale bar: 50μm.

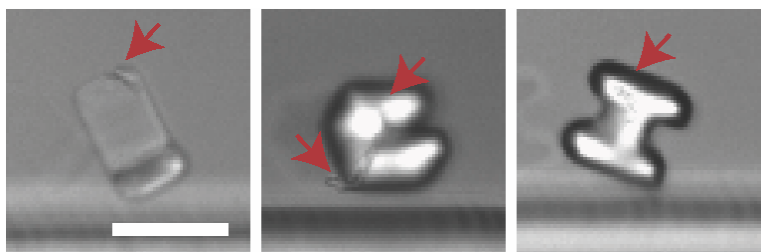

**FIG S 6:** Representative images of observable defects in particles exhibiting inclined orientations. Scale bar: 50μm.

**Table S 1:** Particle feature dimension deviations (see FIG S 3)

| Shape | Log-rolling | | | | Tumbling | | | | Avg CV |
|---|---|---|---|---|---|---|---|---|---|
| | Feature | SD (px) | SD (µm) | CV | Feature | SD (px) | SD (µm) | CV | |
| L | Height | 1.90 | 2.86 | 0.05 | Height | 2.58 | 3.88 | 0.07 | 0.09 |
| | Width | 1.86 | 2.79 | 0.08 | Width | 2.89 | 4.34 | 0.12 | |
| | FeatY1 | 1.90 | 2.86 | 0.05 | FeatY1 | 1.60 | 2.39 | 0.14 | |
| | FeatY2 | 1.83 | 2.75 | 0.15 | | | | | |
| | FeatX1 | 1.90 | 2.86 | 0.05 | | | | | |
| | FeatX2 | 1.83 | 2.75 | 0.15 | | | | | |
| J | Height | 1.75 | 2.63 | 0.05 | Height | 3.27 | 4.91 | 0.09 | 0.13 |
| | Width | 1.97 | 2.95 | 0.06 | Width | 4.95 | 7.43 | 0.18 | |
| | FeatY1 | 1.22 | 1.83 | 0.05 | FeatY1 | 3.04 | 4.56 | 0.26 | |
| | FeatY2 | 1.44 | 2.16 | 0.14 | FeatY2 | 2.25 | 3.37 | 0.22 | |
| | FeatY3 | 1.66 | 2.49 | 0.17 | | | | | |
| | FeatX1 | 1.58 | 2.37 | 0.18 | | | | | |
| | FeatX2 | 1.97 | 2.95 | 0.06 | | | | | |
| | FeatX3 | 1.68 | 2.53 | 0.08 | | | | | |
| V | Height | 1.50 | 2.26 | 0.04 | Height | 2.32 | 3.49 | 0.06 | 0.06 |
| | Width | 1.21 | 1.82 | 0.04 | Width | 3.22 | 4.83 | 0.11 | |
| | FeatY1 | 1.50 | 2.26 | 0.04 | | | | | |
| | FeatY2 | 1.18 | 1.77 | 0.10 | | | | | |
| | FeatX1 | 1.03 | 1.54 | 0.09 | | | | | |
| | FeatX2 | 0.90 | 1.35 | 0.04 | | | | | |
| H | Height | 2.36 | 3.55 | 0.07 | Height | 1.51 | 2.27 | 0.05 | 0.11 |
| | Width | 2.31 | 3.47 | 0.07 | Width | 3.50 | 5.25 | 0.12 | |
| | FeatY1 | 2.36 | 3.55 | 0.07 | FeatY1 | 1.40 | 2.10 | 0.12 | |
| | FeatY2 | 2.36 | 3.55 | 0.07 | FeatY2 | 1.21 | 1.82 | 0.10 | |
| | FeatY3 | 1.65 | 2.47 | 0.16 | | | | | |
| | FeatX1 | 2.04 | 3.05 | 0.17 | | | | | |
| | FeatX2 | 1.79 | 2.69 | 0.15 | | | | | |
| | FeatX3 | 1.80 | 2.70 | 0.17 | | | | | |
| S | Height | 1.79 | 2.69 | 0.05 | Height | 2.77 | 4.16 | 0.07 | 0.16 |
| | Width | 1.74 | 2.61 | 0.05 | Width | 5.46 | 8.19 | 0.21 | |
| | FeatY1 | 1.31 | 1.96 | 0.18 | | | | | |
| | FeatY2 | 1.04 | 1.56 | 0.17 | | | | | |
| | FeatY3 | 1.26 | 1.89 | 0.16 | | | | | |
| | FeatX1 | 2.57 | 3.85 | 0.35 | | | | | |
| | FeatX2 | 1.10 | 1.66 | 0.14 | | | | | |
| U | Height | 1.98 | 2.98 | 0.06 | Height | 1.52 | 2.28 | 0.05 | 0.10 |
| | Width | 1.73 | 2.59 | 0.05 | Width | 2.81 | 4.22 | 0.10 | |
| | FeatY1 | 1.08 | 1.61 | 0.06 | FeatY1 | 1.70 | 2.55 | 0.17 | |
| | FeatY2 | 1.08 | 1.61 | 0.06 | FeatY2 | 1.70 | 2.55 | 0.18 | |
| | FeatY3 | 1.83 | 2.74 | 0.14 | | | | | |
| | FeatX1 | 1.41 | 2.11 | 0.12 | | | | | |
| | FeatX2 | 1.48 | 2.21 | 0.12 | | | | | |
| | FeatX3 | 1.73 | 2.59 | 0.05 | | | | | |

**Table S 2:** MFDs with observable defects in inclined orientations

| Shape | Flow Rate (μl/min) | Clean | Defect | Total | Fraction | Total Defects | Total Counts | Fraction |
|---|---|---|---|---|---|---|---|---|
| L | 370 | 1 | 2 | 3 | 0.67 | 17 | 45 | 0.38 |
| | 916 | 3 | 1 | 4 | 0.25 | | | |
| | 1100 | 4 | 2 | 6 | 0.33 | | | |
| | 1466 | 20 | 12 | 32 | 0.38 | | | |
| V | 370 | 7 | 2 | 9 | 0.22 | 7 | 29 | 0.24 |
| | 916 | 6 | 0 | 6 | 0.00 | | | |
| | 1100 | 6 | 4 | 10 | 0.40 | | | |
| | 1466 | 3 | 1 | 4 | 0.25 | | | |
| J | 370 | 7 | 4 | 11 | 0.36 | 11 | 31 | 0.35 |
| | 916 | 4 | 1 | 5 | 0.20 | | | |
| | 1100 | 1 | 1 | 2 | 0.50 | | | |
| | 1466 | 8 | 5 | 13 | 0.38 | | | |
| S | 370 | 1 | 0 | 1 | 0.00 | 2 | 3 | 0.67 |
| | 916 | 0 | 1 | 1 | 1.00 | | | |
| | 1466 | 0 | 1 | 1 | 1.00 | | | |
| H | 370 | 2 | 4 | 6 | 0.67 | 6 | 15 | 0.40 |
| | 916 | 1 | 0 | 1 | 0.00 | | | |
| | 1100 | 0 | 1 | 1 | 1.00 | | | |
| | 1466 | 6 | 1 | 7 | 0.14 | | | |
| U | 370 | 3 | 3 | 6 | 0.50 | 8 | 22 | 0.36 |
| | 916 | 3 | 2 | 5 | 0.40 | | | |
| | 1100 | 2 | 0 | 2 | 0.00 | | | |
| | 1466 | 6 | 3 | 9 | 0.33 | | | |
| | | | | | | | **Average:** | 0.35 |

## 5. Manual and Statistical Validation of MFD Image Processing Pipeline

To ensure the reliability of both kinematic and geometric measurements, we employed two distinct validation strategies: manual tracings for kinematic features and nonparametric bootstraps for geometric features. First, we compared center-position and angular-displacement outputs (Table S 8a) against manual tracing to gauge tracking accuracy. We manually traced one randomly selected particle from each of the six shapes (L, V, J, S, H, and U) in both log-rolling and tumbling orientations for 10 frames, yielding 240 frames of manual traces in total. The X, Y, and angular-displacement (FIG S 7a) values from manual traces were compared to those from the automated image analysis pipeline using mean absolute difference (MAD) and maximum absolute difference (MAXAD) metrics both in pixels or degrees. Measurement accuracy for particle centers (MAD, FIG S 7) averaged 1 pixel on average in both x and y direction (MAXAD exceeding 2.5 pixels for only 2 of the 28 conditions), and angular displacement errors averaged 4° (MAXAD always less than 15°). These values are well within the expected range, given inherent noise in manual annotation, and demonstrate high accuracy of the image analysis pipeline. Notably the performance of the image analysis code was comparable to state-of-the-art tracking algorithms [8–10]

Second, we assessed the reliability of geometric feature extraction by applying a non-parametric bootstrap to each barcode measurement – computing 95% confidence intervals for height, width, $D_{max}$, and segment-level dimensions (FIG S 3) across flow conditions. The bootstrap method is ideal for this assessment because it makes no assumptions about the data distribution [11,12] and accounts for the observation that faster flows yield more particle observations than slower flows at given recorded duration. We then took the center of each interval as our 'average' feature value, so that our comparisons across different flow rates are both robust and unbiased. Using these barcodes, we found feature-size variations of less than 1-pixel across all flow conditions, confirming that particles - fabricated in a single batch for each shape - were uniform. We then applied the same analysis to compare geometrically analogous dimensions of particles exhibiting log-rolling or tumbling rotation (FIG S 7b); again, differences remain below 1 pixel. Consistency across flow rates (370–1466μL/min) and orientations demonstrates both the reproducibility of our fabrication process and the robustness of our detection pipeline (Table S2-S5). These validations confirm that our dataset is reliable for downstream analysis of hydrodynamic behaviors of inertially focused MFDs.

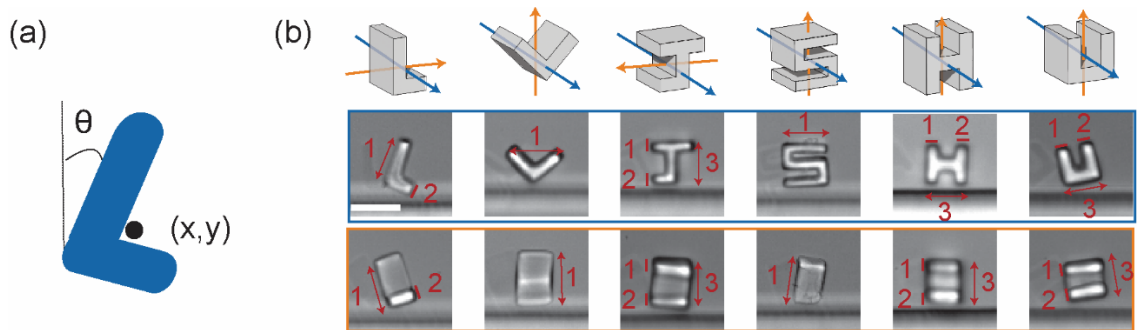

**FIG S 7:** (a) Definition of particle orientation, showing rotation angle ($\theta$) and centroid position (x, y). (b) Comparison of MFD geometries at log-rolling and tumbling orientations. Numbers (1-3) indicate corresponding facet locations on the same particle shape in both orientations. Scale bar: 50μm.

**Table S 3:** Validation of automated tracking accuracy for particle position (X, Y) and orientation ($\theta$), reported as mean and maximum absolute differences (px where 1px=1.5μm).s

|  |  | MAD(X) | MAXAD(X) | MAD(Y) | MAXAD(Y) | MAD($\theta$) | MAXAD($\theta$) |
|---|---|---|---|---|---|---|---|
| L | Log-rolling | 1.44 | 2.74 | 0.80 | 1.54 | 3.62 | 9.99 |
|   | Tumbling | 0.68 | 1.40 | 0.22 | 0.40 | 1.70 | 4.57 |
| J | Log-rolling | 0.69 | 1.30 | 0.83 | 1.47 | 2.83 | 6.12 |
|   | Tumbling | 0.67 | 1.44 | 1.24 | 1.75 | 1.75 | 7.30 |
| V | Log-rolling | 0.82 | 1.93 | 0.65 | 1.17 | 2.00 | 3.98 |
|   | Tumbling | 0.64 | 1.94 | 1.38 | 2.64 | 3.28 | 8.71 |
| H | Log-rolling | 0.75 | 2.03 | 0.47 | 1.26 | 7.38 | 11.46 |
|   | Tumbling | 0.98 | 1.57 | 0.50 | 1.18 | 4.58 | 13.25 |
| S | Log-rolling | 0.81 | 1.74 | 0.80 | 1.54 | 1.38 | 5.55 |
|   | Tumbling | 0.77 | 1.62 | 0.32 | 0.73 | 3.97 | 7.95 |
| U | Log-rolling | 0.56 | 1.05 | 0.65 | 1.21 | 3.50 | 8.22 |
|   | Tumbling | 0.65 | 1.19 | 1.27 | 2.32 | 6.62 | 14.92 |
| + Sq | Log-rolling | 0.54 | 1.80 | 0.31 | 0.71 | 6.33 | 9.82 |
|   | Tumbling | 0.37 | 0.93 | 0.45 | 0.65 | 4.95 | 14.10 |
| **Total** | Log-rolling | 0.75 | 2.74 | 0.62 | 1.54 | 4.00 | 14.10 |
|   | Tumbling | 0.74 | 1.94 | 0.83 | 2.64 | 3.65 | 14.92 |
|   | **Both** | **0.74** | **2.74** | **0.71** | **2.64** | **3.85** | **14.92** |

*Mean Absolute Difference (MAD) and Max Absolute Difference (MAXAD)

**Table S 4:** Comparison of measured particle feature dimensions across videos recorded at different flow rates

| Flow 1 | Flow 2 | Average $\Delta\psi^*$ between Flow 1 and Flow 2 | | | | | | | |
|---|---|---|---|---|---|---|---|---|---|
|   |   | L | H | J | S | V | U | + | Sq |
| 370 | 916 | 0.51 | 0.98 | 0.22 | NaN | 0.10 | 0.42 | NaN | NaN |
| 370 | 1100 | 0.62 | 0.82 | 0.47 | NaN | 0.12 | 0.36 | NaN | NaN |
| 370 | 1466 | 1.00 | 1.05 | 0.17 | NaN | 0.13 | 0.28 | NaN | NaN |
| 916 | 1100 | 0.20 | 0.21 | 0.37 | 0.51 | 0.11 | 0.24 | 0.40 | 0.61 |
| 916 | 1466 | 0.59 | 0.13 | 0.31 | 0.54 | 0.13 | 0.43 | 0.46 | 0.27 |
| 1100 | 1466 | 0.54 | 0.25 | 0.99 | 0.57 | 0.05 | 0.26 | 0.66 | 0.46 |

*$\psi$: Expected value of {h, w, $D_{max}$, Features in X, Features in Y, Comparable Features} for Log Rolling & Tumbling.

**Table S 5:** Comparison of geometric feature dimensions between log-rolling and tumbling orientations

|  | $\Delta\psi^*$ between Log-rolling vs. Tumbling | | |
|---|---|---|---|
| **Shape** | Feature 1 | Feature 2 | Feature 3 |
| L | 0.87 | 1.28 | NA |
| H | 0.67 | 0.15 | 0.02 |
| J | 0.79 | 1.41 | 0.33 |

| | | | | |
|---|---|---|---|---|
| S | 3.71 | NA | NA | |
| V | 5.74 | NA | NA | |
| U | 3.18 | 1.91 | 2.52 | |

*$\psi$=Expected value of features in FIG S 7 for Log Rolling & Tumbling.

**Table S 6:** Assessment of size uniformity in fabricated MFDs based on the coefficient of variation (CV) of height, width and $D_{max}$.

| Shape | Orientation | n | h | w | $D_{max}$ |
|---|---|---|---|---|---|
| L | Log-Rolling | 154 | 0.049 | 0.076 | 0.051 |
| | Tumbling | 90 | 0.065 | 0.119 | 0.073 |
| J | Log-Rolling | 221 | 0.050 | 0.058 | 0.047 |
| | Tumbling | 25 | 0.091 | 0.175 | 0.116 |
| V | Log-Rolling | 407 | 0.044 | 0.036 | 0.034 |
| | Tumbling | 120 | 0.055 | 0.109 | 0.052 |
| H | Log-Rolling | 120 | 0.072 | 0.067 | 0.065 |
| | Tumbling | 7 | 0.045 | 0.119 | 0.063 |
| S | Log-Rolling | 38 | 0.054 | 0.052 | 0.044 |
| | Tumbling | 5 | 0.075 | 0.207 | 0.107 |
| U | Log-Rolling | 2268 | 0.062 | 0.054 | 0.054 |
| | Tumbling | 113 | 0.053 | 0.096 | 0.053 |
| + | Log-Rolling | 405 | 0.112 | 0.098 | 0.098 |
| | Tumbling | 0 | NA | NA | NA |
| Sq | Log-Rolling | 151 | 0.113 | 0.117 | 0.104 |
| | Tumbling | 0 | NA | NA | NA |

## 6. Geometric features of triaxial MFDs

Two-dimensional characteristics were extracted from filled binary masks of each MFD's perimeter, generated by our image analysis pipeline. The rotational diameter, $D_{max}$ (FIG S 8a)– the maximum planar distance between any two perimeter points perpendicular to the rotational axis – and normalized it by the channel width ($D_{max}/W$ [1]). Cross-sectional elongation, $E$ (FIG S 8b), was defined as the ratio of major ($L_a$) to minor ($L_b$) axes of an ellipse fitted to the extremal perimeter points. To compute this, we first extracted those extremal points from the binary mask and centered them by subtracting their mean $x$ and $y$ coordinates. Then, we calculated the covariance matrix of the centered points and performed an eigen decomposition to obtain eigenvalues $\lambda_a$ and $\lambda_b$, which correspond to the squared lengths of the semi-major and semi-minor axes. The full major and minor axes are then $L_a = 2\sqrt{\lambda_a}$ and $L_b = 2\sqrt{\lambda_b}$ respectively, so that $E = \sqrt{\frac{\lambda_a}{\lambda_b}} = \frac{L_a}{L_b}$.

Three-dimensional particle geometries were then reconstructed by taking advantage of the orthogonal "log-rolling" and "tumbling" orientations captured for the same shape-MFDs but in separate frames. From each view, a custom image processing pipeline extracted average cross-sectional shape defined by the corners of the MFDs in the log-rolling orientation (Table S 7) and the extruded depth defined by the width of the tumbling orientation (Table S 8). Corresponding measurements from the two orientations were averaged – assuming sharp edges – to generate a composite 3D model (FIG S 7). From these reconstructions we calculated the Jeffery aspect ratio, $AR = d/D_{max}$ (FIG S 8c), where $d$ is the thickness parallel to the rotational axis [13], volume, $V$, and the moment of inertia, $I$, via voxel summation ($I = \sum m_i r_i^2 = \rho \sum V_i r_i^2$). Here, $\rho$ is the density of MFDs, $V_i$ is volume of one voxel, and $r_i$ is defined as the distance from the rotational axis. Because all PEG hydrogel MFDs were fabricated form the same batch with identical crosslinking method, we assumed they share a uniform density. Thus, we normalized the computed moment of inertial by $\rho$ (i.e., report $I/\rho$) when comparing particles, eliminating any influence of density variations on our geometric analyses. We used the moment of inertia to calculate the diameter of gyration defined as the $D_{Gy} = 2\sqrt{\frac{\rho I}{M}} = 2\sqrt{\frac{I}{V}}$ (FIG S 8d), where $I$ is the rotational moment of inertia, and $M$ is particle mass. In addition to the moment of inertia in log-rolling and tumbling axis of rotation, we calculated the full inertia tensor $I = \begin{bmatrix} I_{xx} & I_{xy} & I_{xz} \\ I_{yx} & I_{yy} & I_{yz} \\ I_{zx} & I_{zy} & I_{zz} \end{bmatrix}$ and diagonalized to yield each MFD's principle moments (FIG S 8e), providing a complete set of three-dimensional descriptors for downstream analysis of MFD hydrodynamics.

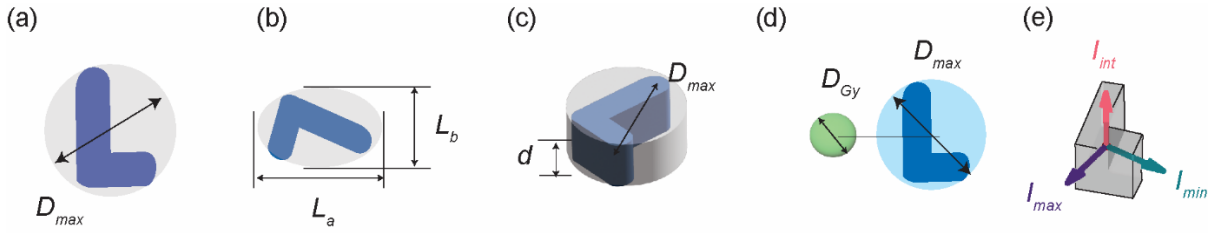

**FIG S 8:** Geometric features of MFDs. (a) Rotational diameter ($D_{max}$). (b) Elongation ($E = \frac{L_a}{L_b}$). (c) aspect ratio ($AR = \frac{d}{D_{max}}$). (d) Diameter of gyration ($D_{Gy} = 2\sqrt{\frac{I}{V}}$). (e) Principal axes of inertia ($I_{min}$, $I_{int}$ and $I_{max}$).

**Table S 7:** Facet coordinates used for 3D reconstruction of MFD geometries (μm).

| Facet ID | L | | H | | J | | S | | V | | U | |
|---|---|---|---|---|---|---|---|---|---|---|---|---|
| | x | y | x | y | x | y | x | y | x | y | x | y |
| 1* | 0 | 0 | 0 | 0 | 0 | 0 | 0 | 0 | 0 | 0 | 0 | 0 |
| 2 | 15 | 0 | 18 | 0 | 51 | 0 | 49.5 | 0 | 18 | 0 | 18 | 0 |
| 3 | 15 | 39 | 18 | 16.5 | 51 | 15 | 49.5 | 10.5 | 18 | 34.5 | 18 | 30 |
| 4 | 36 | 39 | 33 | 16.5 | 31.5 | 15 | 10.5 | 10.5 | 49.5 | 34.5 | 30 | 30 |
| 5 | 36 | 58.5 | 33 | 0 | 31.5 | 52.5 | 10.5 | 19.5 | 49.5 | 51 | 30 | 0 |
| 6 | 0 | 58.5 | 51 | 0 | 0 | 52.5 | 49.5 | 19.5 | 0 | 51 | 48 | 0 |
| 7 | | | 51 | 49.5 | 0 | 37.5 | 49.5 | 49.5 | | | 48 | 48 |
| 8 | | | 33 | 49.5 | 18 | 37.5 | 0 | 49.5 | | | 0 | 48 |
| 9 | | | 33 | 33 | 18 | 15 | 0 | 37.5 | | | | |
| 10 | | | 18 | 33 | 0 | 15 | 39 | 37.5 | | | | |
| 11 | | | 18 | 49.5 | | | 39 | 28.5 | | | | |
| 12 | | | 0 | 49.5 | | | 0 | 28.5 | | | | |

* The top-left corner of each shape set as the origin (0,0).

**Table S 8:** Extruded depth used for 3D reconstruction of MFD geometries.

| Extrusion Depth, $D_{ext}$ (μm) | | | | | |
|---|---|---|---|---|---|
| L | H | J | S | V | U |
| 36 | 43.5 | 42 | 39 | 43.5 | 43.5 |

## 7. Hydrodynamic and Dimensionless Parameters

We characterized particle behavior using two classes of metrics: (1) dimensionless control parameters – the operational Reynolds number, $Re$, and local shear rate, $\dot{\gamma}$, – and (2) kinematic response metrics – lateral equilibrium position, $X_{eq}$, particle rotational angle, $\theta$, rotational velocity, $\omega$, and rotational period, $T$. The local shear rate was defined as $\dot{\gamma} = 2Q/HW^2 = 2U_m/W = 2v/Y$, where $Q$ is the volumetric flow rate, $H$ is the channel height, $W$ is the channel width, $U_m$ is the average fluid velocity, $v$ is the particle velocity, and $Y$ is the particle's center-of-area distance to the nearest channel wall. Because particle velocity may not accurately represent the average fluid velocity in our microchannel due to the particle's focusing position, we replaced the channel width $W$ with the distance from the particle's center of mass to the nearest channel wall, allowing shear rate estimates to more accurately reflect the local shear experienced by the particle. For each MFD, we measured the normalized lateral equilibrium position, $X_{eq} = \frac{Y}{(W/2)}$ (FIG S 4). Particle rotational angle, $\theta$ (FIG S 4), and the rotational period (time for one full revolution), $T$, were obtained by fitting $\theta$ versus time with a linear regression to account for finite frame-rate sampling.

## 8. Statistical Analysis of *Re*-dependent Rotational Orientation Transitions

This section describes the statistical analysis used to generate the Reynolds-number-dependent orientation probabilities shown in FIG 1. We converted each particle's tracked orientation into a binary indicator ($STATE_{LR} = 1$ and $STATE_{TB} = 0$). A Gaussian kernel smoother, a variation of the Nadarya-Watson kernel regression [14], was then applied across $Re$ to estimate the orientation probability, $P(Re)$. The kernel density estimate was computed using a Gaussian kernel defined as $K_h(Re_j, Re_i) = \frac{1}{Z_j} \exp\left(-\frac{(Re_i - Re_j)^2}{2b^2}\right)$ where $Re_j$ is the Reynolds number at which the probability is evaluated, $Re_i$ is the data point from our dataset, and $b$ is the bandwidth. The bandwidth was defined as $b = \frac{Re_{max} - Re_{min}}{4}$, providing sufficient smoothing to reveal overall trends while suppressing noise from uneven sample sizes. The normalized constant, $z_j = \sum_{i=1}^{n} \exp\left(-\frac{(Re_i - Re_j)^2}{2b^2}\right)$, ensures that the estimated probability is normalized to 1. The Gaussian-smoothed proportion of particles in the log-rolling orientation is then given by $f(Re_j) = \sum_{i=1}^{n} K_h(Re_j, Re_i) \cdot STATE_i$.

To evaluate confidence in these estimated proportions, Wald confidence intervals were calculated using the relation, $Z = \sqrt{\frac{n\eta^2}{P(1-P)}}$, where $Z$ is the z-score corresponding to the confidence level, $P$ is the estimated population proportion, and $\eta$ is the desired margin of error. This formulation was used to assess the statistical confidence of the kernel-smoothed orientation curve for the S-MFD, which had a relatively small sample size (n = 43, whereas other MFD shapes had more that 127 and up to 2381 particles per condition; see Table S 6), potentially introducing bias and reducing accuracy. The margin for error was set to $\eta = 0.05$, chosen to match the approximate magnitude of variation observed in orientation proportion, thereby ensuring that uncertainty in the estimates is appropriately captured. This combined approach ensures that the reported $Re$-dependent rotational orientation trends are both smooth and statistically robust.

## 9. Chiral Configuration Effects in Asymmetric MFDs

To investigate whether geometric chirality influences the rotational dynamics of triaxial MFDs, we analyzed in-plane configurations of asymmetric L and J MFDs – each lacking both axial and planar symmetry. Because their log-rolling motion cannot be collapsed into mirrored equivalents - prohibiting direct averaging of rotational speeds - we classified each configuration by chirality. A *chiral* configuration corresponds to a cross-sectional arrangement that matches the readable orientation of the alphabetic shape (i.e., "L" or "J"), while the *achiral* configuration represents its mirror image.

Both L and J MFDs display a slight configuration bias in initial orientation ($Re<50$, FIG S 9). The L-MFD tends to adopt the chiral configuration, whereas the J-MFD more frequently aligns in the achiral configuration. As the Reynolds number increases to ~200, the L-MFD approaches an even distribution between chiral and achiral configuration, while the J-MFD shows a modest increase in the fraction of chiral configurations (FIG S 9). For the L-MFD, this transition coincides with the shift from log-rolling to tumbling orientations (log-rolling proportion <50%).

To understand the geometric origin of these configuration trends, we computed the 2D particle anisotropy ($\alpha$) and triaxiality ($\tau$) from the principal axes of the 2D inertia tensor of the log-rolling orientation (Table S 9). The 2D inertia tensor is defined as $I = \begin{bmatrix} I_{xx} & I_{xy} \\ I_{xy} & I_{yy} \end{bmatrix}$, reflecting the inertia configuration of the particle's cross-sectional shape (e.g. "L" and "J"). The principal moment of inertia are denoted $\varphi_{min}$ and $\varphi_{max}$. Anisotropy is defined as $\alpha = \frac{\varphi_{max}}{\varphi_{min}}$, and triaxiality is defined as $\tau = \sqrt{\left(\frac{\varphi_{min} - \varphi_{avg}}{\varphi_{avg}}\right)^2 + \left(\frac{\varphi_{max} - \varphi_{avg}}{\varphi_{avg}}\right)^2}$, where $\varphi_{avg} = (\varphi_{max} + \varphi_{min})/2$.

The L shape exhibits greater anisotropy and triaxiality than the J shape, indicating a more asymmetric distribution of mass about its centroid. However, because the J MFD has greater overall volume, it also shows a greater absolute difference ($\Delta I$) between its two 2D principal moments. This higher geometric asymmetry (larger $\alpha$ and $\tau$) may contribute to more uneven split between chiral and achiral configurations observed for the J shape at lower Reynolds numbers ($Re\sim50$). Beyond rotational orientation statistics, these chiral configurations also generate distinct instantaneous drag-area profiles during rotation, motivating further analysis of drag effects on particle dynamics. (Supplementary Section 10).

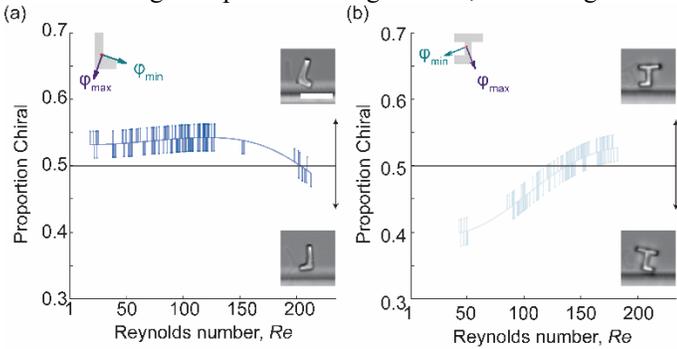

**FIG S 9:** Proportion of particles in chiral orientation for (a) L-MFD (b) J-MFD. Scale bar: 50μm.

Table S 9: Principal axes of the 2D inertia tensor

| Shape | 2D $I$ ($10^{-22}$ μm$^5$) | | $\Delta I$ | Anisotropy, $\alpha$ | Relative Difference, $\frac{|\varphi_{max} - \varphi_{min}|}{\varphi_{max}}$ | Triaxiality, $\tau$ |
|---|---|---|---|---|---|---|
| | $\varphi_{min}$ | $\varphi_{max}$ | | | | |
| L | 0.03 | 0.15 | 0.11 | 4.59 | 0.78 | 0.91 |
| J | 0.08 | 0.21 | 0.12 | 2.51 | 0.60 | 0.61 |

## 10. Drag Area Effects on Particle Rotational Motion

Building on the chirality-dependent configuration analysis, we next examined how variations in projected drag area influence the rotational dynamics of MFDs. Jeffery's classical theory describes the rotational dynamics of ellipsoids in shear flow as a function of aspect ratio and shear rate [13]. Subsequent extensions incorporated cylindrical particles by modifying the aspect ratio as $AR = c \cdot AR_{measured}$ [15,16], where $c$ accounts for differences in projected drag area [17], highlighting the critical role of shape-dependent drag in setting rotational behavior. Leveraging 3D reconstructions of our MFDs, we computed orientation-dependent drag areas and compared them to the measured, normalized rotation rates and angles. For log-rolling MFDs, drag-area oscillations were in phase with angular velocity—higher angular velocity coincided with greater drag area—across all shapes (FIG S 10-11). Notably, the chiral J and L particles exhibited mirrored instantaneous dynamics in both angular velocity and drag area, despite having similar average behavior (FIG S 10). This mirroring arises from the opposite alignment of surface facets and internal voids relative to the flow direction, producing inverted drag-area profiles between the two chiral configurations. In contrast, tumbling MFDs showed a phase mismatch between angular velocity and drag area, likely due to the alignment of internal voids along the drag axis, altering the effective drag during rotation (FIG S 12). These findings provide the basis for extending Jeffery's model to account for geometry-dependent inertial effects.

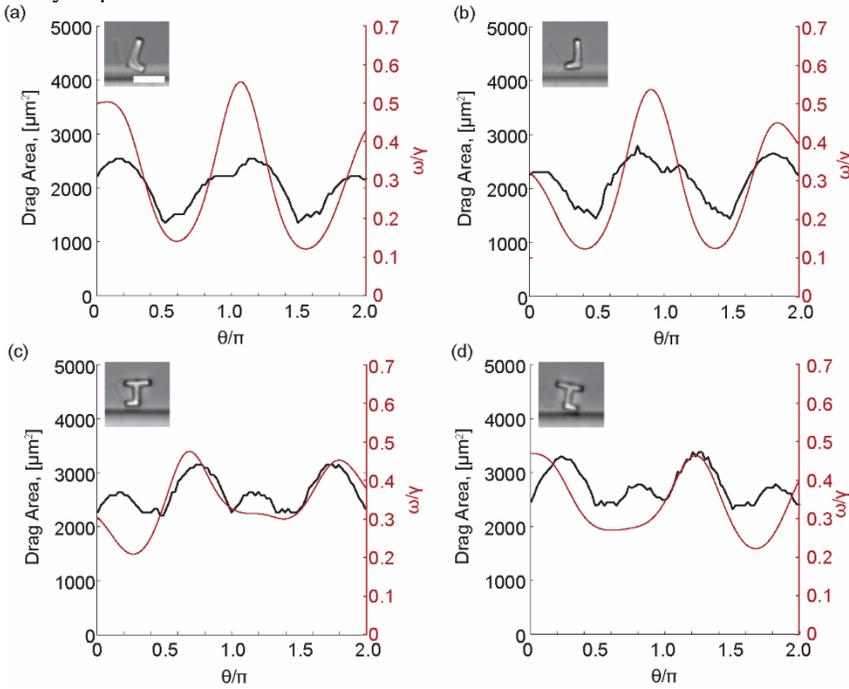

**FIG S 10:** Drag area and normalized angular velocity as a function of orientational angle ($\theta/\pi$) for log-rolling L- and J-shaped MFDs in chiral and achiral configurations. (a and b) L-MFDs in chiral and achiral configurations, respectively. (c and d) J-MFDs in chiral and achiral configurations, respectively. Black curves represent the projected drag area, and red curves show the angular velocity normalized by the local shear rate ($\omega/\dot{\gamma}$). The orientational angle ($\theta/\pi$) is normalized to represent one complete rotation cycle. Insets illustrate corresponding particle configurations within the flow. Scale bar: 50µm. Rotation data ($\theta/\pi$, $\omega/\dot{\gamma}$) represent smoothed averages over n = 84 (L-chiral), 70 (L-achiral), 99 (J-chiral), and 122 (J-achiral) particles. The drag area is calculated by rotating the 3D rendering according to the orientational angle derived from the smoothed average.

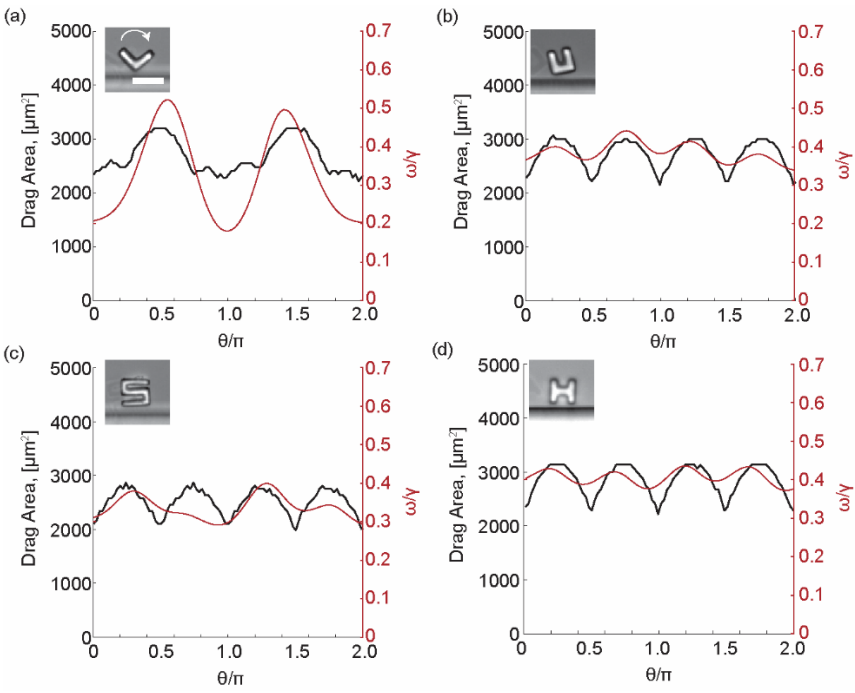

**FIG S 11:** Drag area and normalized angular velocity as a function of orientational angle ($\theta/\pi$) for log-rolling MFDs of various shapes. (a) V-shaped, (b) U-shaped, (c) S-shaped, and (d) H-shaped MFDs. Black curves represent the projected drag area, and red curves show the angular velocity normalized by the local shear rate ($\omega/\dot{\gamma}$). The orientational angle ($\theta/\pi$) is normalized to represent one complete rotation cycle. Insets illustrate corresponding particle configurations within the flow. Scale bar: 50μm. Rotation data ($\theta/\pi$, $\omega/\dot{\gamma}$) represent smoothed averages over n = 407 (V), 2268 (U), 38 (S), and 120 (H) particles. The drag area is calculated by rotating the 3D rendering according to the orientational angle derived from the smoothed average.

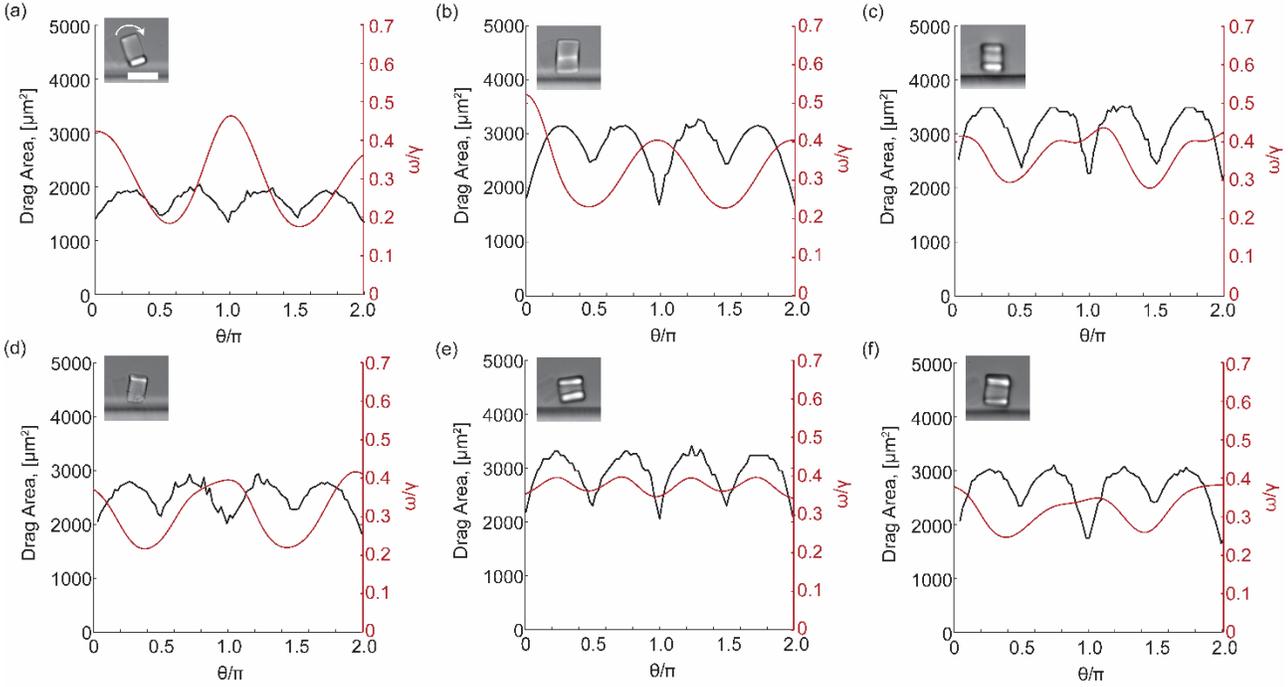

**FIG S 12:** Drag area and normalized angular velocity as a function of orientational angle ($\theta/\pi$) for tumbling MFDs. (a) L-shaped, (b) V-shaped, (c) J-shaped, (d) S-shaped, (e) U-shaped, and (f) H-shaped MFDs. Black curves represent the projected drag area, and red curves show the angular velocity normalized by the local shear rate ($\omega/\dot{\gamma}$). The orientational angle ($\theta/\pi$) is normalized to represent one complete rotation cycle. Insets illustrate corresponding particle configurations within the flow. Scale bar: 50μm. Rotation data ($\theta/\pi$, $\omega/\dot{\gamma}$) represent smoothed averages over n = 90 (L), 120 (V), 25 (J), 5 (S), 7 (H), and 113 (U) particles. The drag area is calculated by rotating the 3D rendering according to the orientational angle derived from the smoothed average.

## 11. Jeffery Orbit Equation of Motion

Extending the insights from the drag-area analysis, we evaluated whether Jeffery's classical framework can accurately capture the rotational dynamics of MFDS. In creeping flow, triaxial particles can exhibit chaotic or double-periodic rotations [1,18–22]; however, in our inertial Poiseuille experiments, MFDs stabilize into two Jeffery-like modes: log-rolling and tumbling. To test whether Jeffery's classical period, $T =$

$\left(AR + \frac{1}{AR}\right)\left(\frac{2\pi}{\dot{\gamma}}\right)$ [13], applies, we measured rotation period of each MFD and compared it to the theoretical prediction. Here, $AR = d/D_{max}$ is the aspect ratio of MFDs, and the local shear rate is defined as $\dot{\gamma} = 2v/Y$, where $v$ is the particle velocity and $Y$ is the distance between the particle centroid and the nearest channel wall, both obtained from high-speed imaging.

Jeffery's formula consistently overestimated the measured periods, with the magnitude of deviation varying systematically with shape and rotational orientation (FIG S 13). Previous cylinder-based corrections accounted for geometric drag by introducing an empirical factor $c$ to modify the aspect ratio ($AR$ ($AR = c \cdot AR_{Measured}$), where $c = 0.7$ for prolate rods [15,17,23] and $1.6 < c < 2.3$ for oblate disks [24]. Unlike previous cylinder-based models, applying a single scalar correction to $AR$ did not reconcile the predicted and measured periods for MFDs, indicating that such simplifications fail to capture the influence of multifaceted geometry. This systematic variation suggests that factors beyond aspect ratio – specifically, the particle's mass distribution relative to its rotation axis- strongly influence the rotational period, since the moment of inertia quantifies a body's resistance to rotation for a given mass.

Although all MFDs share an oblate morphology ($AR \approx 0.6 \pm 0.1$) and exhibited Jeffery-type periodic oscillations in angular velocity (FIG S10-12), they still display strong shape- and orientation-dependent deviations from Jeffery's predicted periods. These deviations arise from both shape-dependent mass distribution and local drag fluctuations, which must be incorporated into any predictive model. The U-MFD follows a nearly sinusoidal, constant-amplitude $\omega/\dot{\gamma}$, whereas the L-MFD shows a skewed, non-sinusoidal profile with sharp peaks and deep troughs as its facets alternately align and misalign with the flow, clearly illustrating how cavities and nonuniform edges modulate the instantaneous rotation rate. These asymmetries demonstrate that both shape-dependent mass distribution and local drag fluctuations must be built into any predictive model.

To account for these asymmetries, we formulated an inertia-adapted Jeffery equation: $T = C_m \frac{I_o}{I_{rot}}\left(AR + \frac{1}{AR}\right)\left(\frac{2\pi}{\dot{\gamma}}\right)$ where $C_m = 2$ is an empirically determined scaling factor, $I_{rot}$ is the moment of inertia of the MFD about its active rotation axis, and $I_o = \frac{1}{12}M(h^2 + w^2) = \rho \frac{1}{12} hwd (h^2 + w^2)$ is the moment of inertia of a bounding rectangular prism with the same height ($h$), width ($w$), and depth ($d$).

This model achieved strong agreement with experiments ($R^2 > 0.8$; average $R^2_{ave} = 0.89$) across all cases except for the S-tumbling and J-shaped MFDs (Table S10). The S-tumbling case showed lower fit quality ($R^2 = 0.45$) due to a small sample size (n=5), while J-MFDs exhibited $R^2$ values of 0.40 for log-rolling and 0.77 for tumbling, primarily reflecting fabrication inconsistencies. Notably, the coefficient of variation (CV) for feature lengths in J-MFDs averaged 13%, with some instances reaching up to 26% (Table S1), compared to an average CV of 11% for all other samples. Because the moment of inertia scales with the fifth power of the characteristic dimension, even modest geometric variability can substantially impact predictive accuracy. Despite these limitations, the strong consistency observed across multiple shapes and orientations confirms that incorporating the geometry-dependent inertia ratio ($C_m \frac{I_o}{I_{rot}}$) substantially enhances predictive accuracy relative to the classical Jeffery model. Further discussion of this enhancement is provided in the main text.

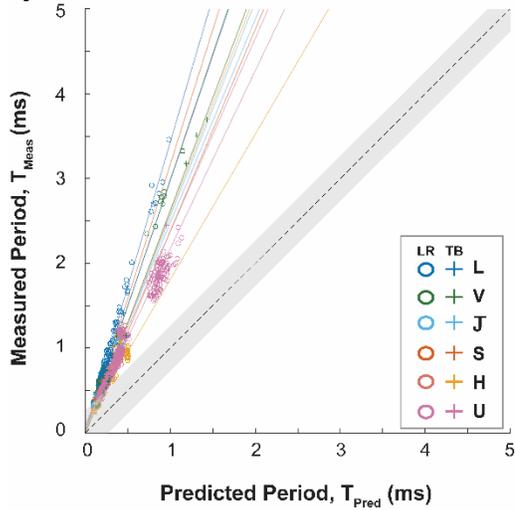

**FIG S 13:** Comparison of measured and predicted rotational periods ($T_{meas}$ vs. $T_{Pred}$) for triaxial MFDs using Jeffery's original model. triaxial multifaceted disks. The 95% confidence interval (shaded region) highlights systematic overestimation of rotational periods by the original Jeffery model. Data correspond to the analysis used in FIG 2.

Table S 10: Validation of the inertial-adapted Jeffery orbit equation across MFD shapes and orientations.

| Shape | Orientation | $I_o$ (μm⁵) | $I_{rot}$ (μm⁵) | AR | n | $R^2$ |
|---|---|---|---|---|---|---|
| L | Log-Rolling | 5.96E-23 | 1.77E-23 | 0.532 | 154 | 0.99 |
| | Tumbling | 5.96E-23 | 2.01E-23 | 0.524 | 90 | 0.96 |
| V | Log-Rolling | 9.25E-23 | 2.52E-23 | 0.614 | 407 | 0.87 |
| | Tumbling | 9.25E-23 | 2.85E-23 | 0.465 | 120 | 0.86 |
| J | Log-Rolling | 1.00E-22 | 2.90E-23 | 0.578 | 221 | 0.40 |
| | Tumbling | 8.47E-23 | 2.93E-23 | 0.746 | 25 | 0.77 |
| S | Log-Rolling | 7.80E-23 | 3.23E-23 | 0.561 | 38 | 0.95 |
| | Tumbling | 6.32E-23 | 2.39E-23 | 0.725 | 5 | 0.45 |
| H | Log-Rolling | 9.25E-23 | 3.94E-23 | 0.615 | 120 | 0.80 |
| | Tumbling | 8.22E-23 | 3.73E-23 | 0.739 | 7 | 0.86 |
| U | Log-Rolling | 7.70E-23 | 3.56E-23 | 0.642 | 2268 | 0.93 |
| | Tumbling | 7.01E-23 | 3.24E-23 | 0.782 | 113 | 0.81 |

## 12. Additional Square-Ring and Plus Geometries for Inertial Focusing Analysis

Although all MFDs were designed with similar rotational diameters and aspect ratios, they exhibited varying degrees of cross-sectional elongation and symmetry. To assess whether the observed lateral equilibrium trends extended beyond these six shapes, we introduced two additional symmetric geometries—a plus-shaped (+) and a square-ring (Sq) disk—with smaller rotational diameters ($D_{max}/W \sim$ 0.4 and 0.5, respectively) and thinner profiles ($AR \approx 0.25$). These two shapes were used exclusively for lateral equilibrium analysis and were excluded from other shape-dependent comparisons.

Several key differences emerged in the inertial focusing behavior of the plus and square-ring MFDs. While both exhibited log-rolling dynamics, neither transitioned to a feature-aligned tumbling orientation (i.e., a 90° offset from log-rolling). Instead, they adopted either inclined or atypical rotational states. The inclined state closely resembled log-rolling in that rotation occurred about a particle symmetry axis, but with that axis tilted away from the channel's vorticity direction. This behavior is consistent with previous reports of inclined rolling in symmetric disk-shaped particles in inertial Poiseuille flow [25], suggesting that both plus and square-ring MFDs possess sufficient symmetry to suppress multifacetedness-induced tumbling.

The atypical state was characterized by particles intermittently leaving the imaging focal plane, often accompanied by blurred particle edges. We hypothesize that this behavior correspond to inclined particles temporarily moving out of focus due to their smaller rotational diameters. Supporting this interpretation, we observed instances of plus-shaped particles adopting atypical orientations centered within the channel (FIG S 14d gray), suggesting the presence of four lateral focusing positions reminiscent of focusing behavior in square-cross-section channels [26].

Because plus and square-ring MFDs did not exhibit tumbling orientations, their aspect ratios could not be estimated using the 3D reconstruction methods described above. However, we found that the facet length in the log-rolling state ($F_a$) closely matched the apparent extrusion depth ($d$) in inclined orientations (FIG S 14a). By assuming $F_a \approx d$, we estimated their aspect ratios as $AR=d/D_{max} \sim 0.25$, significantly lower than those of other MFDs. This reduced $AR$, combined with their higher geometric symmetry, likely favors inclined rolling akin to that of axisymmetric cylindrical disks rather than shape-induced tumbling.

Finally, we found that the square-ring shape exhibited log-rolling behavior in only 40% of cases, compared to 93% for the plus shape. The central void in the square-ring geometry likely introduces additional hydrodynamic effects that alter both rotational dynamics and lateral focusing behavior [27,28], consistent with the distinct equilibrium positions observed for these shapes in our main results. These observations highlight that geometry symmetry and reduced aspect ratio together favor inclined rolling and suppress shape-induced tumbling in MFDs.

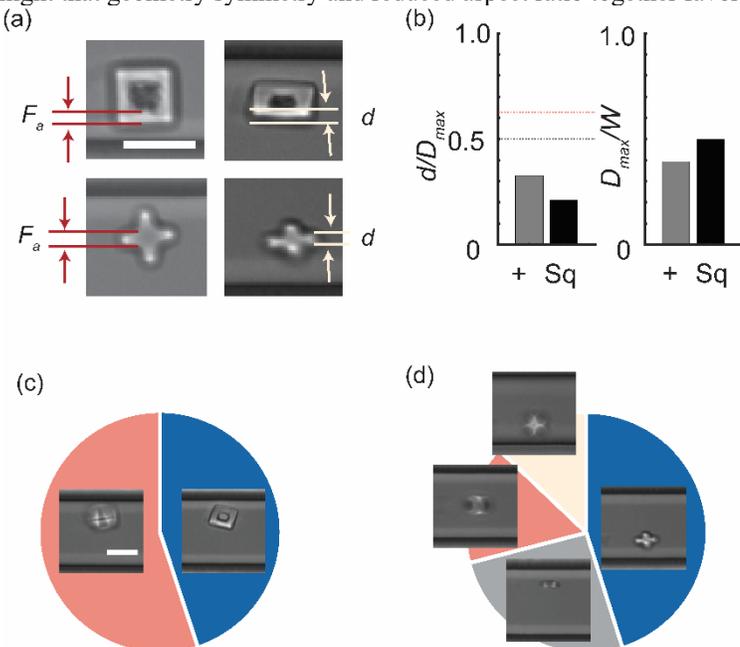

**FIG S 14: Square-ring (Sq) and** plus (+) geometries used for inertial focusing analysis. (a) Features observed in the log-rolling orientation used to estimate the extruded depth ($d$). The facet length ($F_a$) was approximately equal to the extruded depth for both square-ring and plus shapes. (b) Aspect ratio ($AR=d/D_{max}$) and rotational diameter ($D_{max}/W$) for plus and square-ring MFDs, with red lines indicating the average values of $AR$ and $D_{max}/W$ for other MFDs. Fraction of particles exhibiting inclined (blue) and atypical (orange, gray, yellow) orientations for (c) square-ring and (d) plus-shaped MFDs. Scale bar: 50μm.

Table S 11: Inclined vs. Atypical Counts for square-ring and plus MFDs

|  | Log-rolling | Inclined | Atypical | Total | % of LR |
|---|---|---|---|---|---|
| Sq | 151 | 102 | 125 | 378 | 40 |
| + | 405 | 14 | 17 | 436 | 93 |

## II.  SUPPLEMENTARY VIDEOS

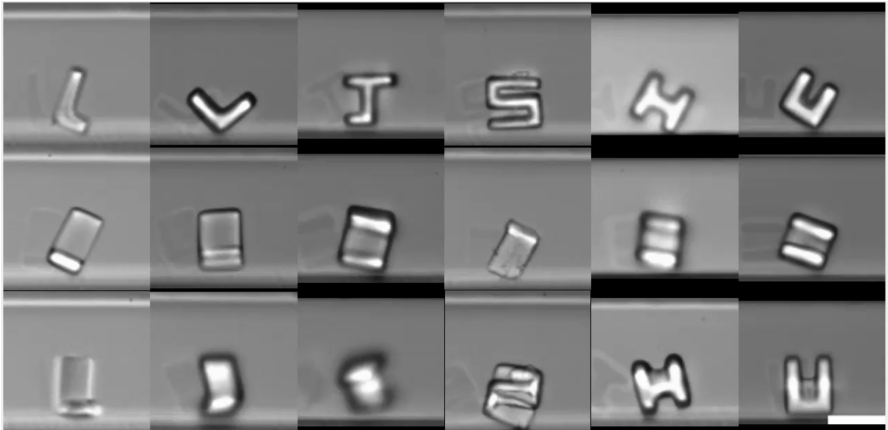

**SI Movie 1:** Still images from high-speed video showing the rotational motion of multifaceted disks. The first, second and third rows correspond to log rolling, tumbling, and inclined orientations, respectively. Scale bar: 50μm.